\documentclass[letterpaper,journal,final]{IEEEtran}

\usepackage{amssymb}
\usepackage{amsmath}
\usepackage{amsfonts}                                
\usepackage[latin1] {inputenc}
\usepackage{enumerate}
\usepackage{mathrsfs}
\usepackage{tabulary}
\usepackage{cite}
\usepackage{psfrag}
\usepackage{cite}
\usepackage{graphicx}          
\usepackage{color}      
\usepackage{epsfig} 
\usepackage{epstopdf}
\usepackage{times} 
\def\qedp{\hspace*{\fill}~{\tiny $\blacksquare$}}
\def\qed{\relax\ifmmode\hskip2em \Box\else\unskip\nobreak\hskip1em $\Box$\fi}

\newtheorem{theorem}{Theorem}
\newtheorem{itlemma}{Lemma}
\newtheorem{itdefinition}{Definition}
\newtheorem{itproposition}{Proposition}
\newtheorem{itresult}{Result}
\newtheorem{itremark}{Remark}
\newtheorem{itassumption}{Assumption}
\newtheorem{itcorollary}{Corollary}
\newtheorem{itexample}{Example}

\newenvironment{remark}{\begin{itremark}\rm}{\end{itremark}}

\newenvironment{corollary}{\begin{itcorollary}\rm}{\end{itcorollary}}

\title{\LARGE An Integrated Human-physical Framework for Control of Power Grids } 

\author{S. Feng, M. Cucuzzella, T. Bouman, L. Steg and J. M. A. Scherpen

	\thanks{Corresponding author: J. M. A. Scherpen.
		 The material in this paper was not presented at any conference.}
\thanks{Shuai Feng, Michele Cucuzzella and Jacquelien Scherpen
	 are with ENTEG, Faculty of Science and Engineering, University of Groningen, 9747AG Groningen, the Netherlands ({\tt\small  s.feng@rug.nl, m.cucuzzella@rug.nl, j.m.a.scherpen@rug.nl})
	 . }
	\thanks{	Thijs Bouman and Linda Steg are with Faculty of Behavioural and Social Sciences, University of Groningen, Grote Kruisstraat 2/1, 9712TS Groningen, the Netherlands
		 ({\tt\small t.bouman@rug.nl,  e.m.steg@rug.nl}).}
	 	
		\thanks{This research is supported by the incentives and algorithms for efficient,
			reliable, sustainable and socially acceptable energy system integration
			(ERSAS) research program funded by the Dutch organization for scientific
			research (NWO) with grant number 647-002-005.}
}

\IEEEoverridecommandlockouts  
\begin{document}
\maketitle
\begin{abstract}
In this paper, we bridge two disciplines: systems \& control and environmental psychology. We develop second order Behavior and Personal norm (BP) based models (which are consistent with some studies on opinion dynamics) for describing and predicting human activities related to the final use of energy, where psychological variables, financial incentives and social interactions are considered. Based on these models, we develop a human-physical system (HPS) framework consisting of three layers: (i) human behavior, (ii) personal norms and (iii) the physical system (i.e., an AC power grid). Then, we formulate a social-physical welfare optimization problem and solve it by designing a primal-dual controller, which generates the optimal incentives to humans and the control inputs to the power grid. Finally, we assess in simulation the proposed models and approaches.

\end{abstract}


%
%
%
%
%
\section{Introduction}
Individuals' behavior is critical for the functioning of energy systems. Accordingly, understanding the drivers behind this behavior is therefore key for the modeling and optimization of energy systems. For example, knowledge on such drivers could be employed to better understand individuals' energy behavior that affects the energy system's functioning, and to promote the behavior that makes the energy system function more optimally, possibly enhancing the effectiveness of technical solutions\cite{steg2015understanding, stern2016opportunities}.
For better understanding human behavior and its impact on physical systems, one may need to develop mathematical models from ``a control perspective" that capture drivers of energy behavior. 

This paper presents interdisciplinary work integrating systems \& control and environmental psychology.
We first develop feasible mathematical models of dynamical human behavior. Then, to model the impact of dynamical human behavior on a power grid,  
to further model the effects of incentives on energy use behavior, and to
 obtain the control inputs to the power grid as well as the incentives to humans, we develop a human-physical system (HPS) framework. Specifically, the HPS includes three layers: a behavior layer and a personal norm layer which describe social human activities, and a physical layer that describes the dynamics of an AC power grid.

\subsection{Energy saving behavior in environmental psychology}
The dynamical human behavior models proposed in this paper are inspired by and consistent with findings in psychology.
Various studies have examined which factors affect energy use behavior\cite{steg2015understanding, stern2016opportunities, stern2016towards, sovacool2014diversity}. It is well established that energy use behavior is rooted in an individual's \emph{personal values} (values for short), which reflect general goals that people strive for in their life. Personal values typically influence an individual's pro-environmental behavior via \emph{personal norms}, reflecting feelings of moral obligation to act pro-environmentally \cite{schwartz1981normative,stren2000toward,stern1999value}.

When focusing on energy saving behavior, which is central in this paper, two values appear particularly relevant: \emph{egoistic values} that reflect a concern with self interest, status and possessions, and \emph{biospheric values} that reflect a concern to protect and care about nature and the environment\cite{steg2015understanding, dietz2005environmental, steg2018drives}. 
Previous research indicated that stronger endorsement of biospheric values likely results in stronger personal norms to save energy, which in turn increase the likelihood that someone will actually engage in energy saving behavior.
Conversely, stronger endorsement of egoistic values is typically indicative of weaker personal norms to save energy\cite{schwartz2015advertising}, which may de-motivate energy saving behavior, although this effect is relatively weak.
In addition to values' influence on energy behavior via personal norms, individuals may also be motivated to reduce their energy consumption because of associated financial benefits (e.g., lower energy bills). Particularly individuals with stronger egoistic values may save energy because of such financial benefits, and those individuals may be particularly motivated by \emph{financial incentives} (incentives for short), e.g., subsidies. 
We refer the readers to Section II-B for the detailed explanations of personal values, egoistic and biospheric values, personal norms and financial incentives. 
Moreover, social influence can also affect people's energy consumption. For instance, social norms (i.e., perceptions that others save energy or expect you to save energy) can motivate people to engage in sustainable energy saving behavior\cite{nolan2008normative}, provided that the majority does engage in such behavior \cite{schultz2007constructive}. 

\subsection{Human-physical system (HPS) framework}

In view of the last subsection, this paper develops human activity models of energy saving incorporating 1) behavior that depends on financial incentives; 2) personal norms that are influenced by egoistic and biospheric values, and social norms. The interactions between these items are shown in Figure \ref{interfig}. Item 1) is considered to be the behavior layer whose response is generally quick, while item 2) is considered to be the personal norm layer whose response is generally slower than the one of the behavior layer.
Specifically, energy saving behavior is influenced by the underlying personal norms and financial incentives. From a control perspective, the incentives can be considered as ``control inputs" (i.e., behavior interventions) to the behavior dynamics, while 
personal norms can be considered as references for the behavior in absence of incentives. From an opinion dynamics viewpoint, personal norms can be considered as the opinions of the individuals in a social network. 
It is worth mentioning that our models of describing human activities in energy saving behavior are partially inspired by and consistent with studies in opinion dynamics (see for instance the continuous-time Friedkin-Johnsen and high-order opinion dynamics models 
\cite{degroot1974reaching, friedkin1990social, ye2020continuous, wongkaew2015control, caponigro2016sparse, amelkin2017polar}).


For the physical layer, we consider an AC power grid where people (prosumers) share the task of current generation (current sharing) with peers in their local electricity network according to their generation
capacities, regulating the voltage within permitted (safe) limits around the desired value (voltage regulation).
More precisely, current sharing and voltage regulation are vital control objectives for preventing energy source overstressing and guaranteeing stability, respectively\cite{nasirian2014distributed, de2018power, cucuzzella2018robust, trip2018distributed}. 
Our work is inspired by the recent paper \cite{cucuzzella2019distributed}, which studies a social-physical welfare optimization
problem depending on  prosumers' motives. We will clarify in the next subsection the contributions of our work and the differences with respect to \cite{cucuzzella2019distributed}. 

\subsection{Contributions}

We formulate a convex social-physical welfare optimization problem, whose solution corresponds to ``control inputs" to both humans (i.e., behavior interventions such as financial incentives) and to the power grid. Specifically, we aim at i) maximizing the social welfare by satisfying the prosumers' load demand and minimizing the incentives' cost, and ii) maximizing the  physical welfare by performing current sharing and voltage regulation. To achieve these goals, we design a dynamic controller, whose unforced dynamics represent the primal-dual dynamics of the considered optimization problem \cite{cucuzzella2019distributed, stegink2016unifying, kosaraju2019distributed, zhao2014design }. The contributions of this paper are four-fold:
\begin{itemize}
	\item We bridge systems \& control and environmental psychology, by developing a novel HPS framework to study and analyze energy saving dynamical behavior of humans taking into account the dynamics of the physical infrastructures of a power grid. To the best of our knowledge, such a framework is completely novel and has never been developed or studied from a control perspective. More precisely, we develop second-order models describing Behavior and Personal norm (BP) dynamics, which include intrinsic (i.e., values and personal norms) and extrinsic drivers of energy saving behavior, allowing to model their slow and fast transient processes, respectively. These are significant improvements with respect to \cite{cucuzzella2019distributed}, where the prosumers are supposed to have non-dynamical behavior and hence no dynamic models describing human activities are developed.
	\item Inspired by the findings in environmental psychology, this paper considers a scenario where values, incentives and social influence affect prosumers' energy saving behavior instead of letting automation directly adjust prosumers' load, as supposed in \cite{cucuzzella2019distributed}. Clearly, this more realistic scenario where the prosumers' load demand depends on dynamical human behavior makes the theoretical analysis different and more complex than the one in \cite{cucuzzella2019distributed}. 	
	\item We consider an AC power grid as physical layer, whose dynamics are more complex than (and include) the DC counterpart studied in \cite{cucuzzella2019distributed}. 
	\item Differently from \cite{cucuzzella2019distributed}, where the controller generates inputs only to the power grid, the proposed HPS framework and primal-dual controller provide also optimal and socially acceptable incentives to humans, affecting their energy saving behavior. 
\end{itemize}

\textbf{Outline.} This paper is organized as follows. In Section II, we develop the overall HPS framework including the power grid and Behavior and Personal norm models for human activities. In Section III we present the control objectives and in Section IV we design the primal-dual controller and analyze the closed-loop stability.  
A numerical example is presented in Section V, and finally Section VI ends the paper with conclusions and future research.

\textbf{Notation.} We denote by  $\mathbb R$ the set of reals. Given $y \in \mathbb R$, $\mathbb R_{\geq y}$ denotes the set of reals no smaller than $y$.
For any $w \in \mathbb Z$, we denote $\mathbb Z_{\ge w} := \{w,w+ 1,\cdots\}$. Let $\mathbf 0$ and $\mathbf 1$ denote column vectors of appropriate dimensions, having all $0$ and $1$ elements, respectively. Let \textbf{I} denote the identity matrix with appropriate dimension. 
Given a vector $v$, let $\|v\|$ denote its $\ell_2$ norm. 
We let $\mathcal N $ denote the set of $N \in \mathbb Z_{\ge 2} $ prosumers and $\mathcal E  $ denotes the set of $E  \in \mathbb Z_{\ge 1} $ transmission lines interconnecting the prosumers. Moreover let $\mathcal N_i  \subseteq \mathcal N$ denote the set of prosumers physically interconnected with prosumer $i$ in the power grid, and $\mathcal E _i \subseteq \mathcal E$ denote the set of the transmission lines connected to prosumer~$i$. Let $\mathcal S_i  \subseteq \mathcal N$ denote the set of the social neighbors of prosumer $i$.  

\section{Human-physical system (HPS) framework}

For the readers' convenience, before introducing the overall HPS framework, we will first introduce the AC power grid model and then the Behavior and Personal norm (BP) models.

\subsection{AC microgrid model}

\begin{table}[t]
	\centering
	\caption{Parameters and Variables  ($g=d,q$)}
	\begin{tabular}{ l l l l}
		\hline
		\multicolumn{2}{c}{Table of parameters}  \\ 
		\hline
		$L_{ti}$ & Filter inductance \\
		$R_{ti}$ & Filter resistance   \\
		$L_k$ & Line inductance      \\
		$R_k$ & Line resistance   \\
		$C_{ti}$ & Filter capacitance   \\
		$R_{Li}$ & Load impedance  \\
		$I_{Lgi}$ & Load current\\
		\hline
	\end{tabular}
	\begin{tabular}{ l l l l}
		\hline
		\multicolumn{2}{c}{Table of variables }  \\ 
		\hline
		$I_{tgi}$ & Generated current \\
		$V_{gi}$ & Load voltage \\
		$I_{gk}$ & Line current\\
		$u_{gi}  $ & Control input \\
		$z_{li}$ & Energy behavior  \\
		$p_i$ & Personal norms\\
		$s_i$ & Incentives\\
		\hline
	\end{tabular}
	\label{table 1}

\end{table}

In this paper, we consider a low-voltage AC power grid composed of $N $ prosumers that are connected by $E $ resistive-inductive transmission lines. 
From a physical point of view, we assume that every prosumer can be represented by a distributed generation unit including a Voltage
Sourced Converter (VSC) and a load.
Moreover, we recall that in low-voltage grids the lines are predominately
resistive, leading to a strong coupling between the active power flows
and the voltage magnitude \cite{guerrero2007decentralized,golsorkhi2016gps}. Also, we assume that the frequency 
 is controlled in open-loop by equipping each VSC
with an internal oscillator that provides the phase angle $\delta(t) = \int_{0}^{t} \omega_0 dt$, with $\omega_0 = 2\pi f_0$ and $f_0$ denoting the nominal frequency of the grid. Then, provided that the power grid is
balanced and symmetric, and all clocks of the internal oscillators are synchronized, we apply Clarke's and Park's transformation to obtain the system dynamics in the rotating $dq$-frame\cite{park1929two}.

Given the notation in Table \ref{table 1}, 
the dynamics of the physical system at prosumer $i$ can be expressed as in \cite{cucuzzella2018cooperative}, i.e.,
\begin{subequations}\label{physical i}
\begin{align}
C_{ti} \dot V_{di} &=  \omega_0 C_{ti}  V_{qi} + I_{tdi} + \sum_{k \in \mathcal E _ i} I_{dk}  -\frac{V_{di}}{R_{Li}} -     I _{Ldi} z_{li} \label{V di}       \\
C_{ti} \dot V_{qi} &=  -  \omega_0 C_{ti}  V_{di} + I_{tqi} + \sum_{k \in \mathcal E _ i} I_{qk} - \frac{ V_{qi} }{R_{Li} }- I _{Lqi} z_{li} \label{V qi}     \\
L_{ti} \dot I_{tdi} &=  - V_{di} - R_{ti} I_{tdi} + \omega_0 L_{ti} I_{tqi} + u_{di} \label{I tdi} \\
L_{ti} \dot I_{tqi} &=  - V_{qi} - R_{ti} I_{tqi} - \omega_0 L_{ti} I_{tdi} + u_{qi}  \label{I tqi}, 
\end{align}
\end{subequations}
where the $d$ and $q$ subscript represent the direct and quadrature component, respectively. For instance, $V_d$ and $V_q$ are the
$d$ and $q$ components of voltage, respectively.
The resistance $R_{Li}$ can represent the base impedance load of prosumer $i$ or simply the system damping,  while $I _{Lgi} z_{li}$  ($g=d,q$) represents the dynamic current load. 
In particular, $I_{Ldi}$ and $I_{Lqi}$ are constants that represent the load demand of prosumer $i$, while 
 $z_{li}: \mathbb R_{\geq 0} \rightarrow [0,1]$ is a dynamical variable depending on the behavior of prosumer $i$. 
For example, if $z_{li}=0$ or $z_{li}=1$ for all $t$, then the actual load of prosumer $i$ is $0$ or $I_{Lgi}$, respectively. The dynamics of the behavior $z_{li}$ will be further explained in the next subsection. 

In (\ref{physical i}), $I_{dk}$ and $I _ {qk} $ denote the current exchanged between prosumers $i$ and $j \in  \mathcal N_i$ through the line $k\in \mathcal{E}_i$.  
The ends of the transmission line connecting prosumers $i$ and $j$ are arbitrarily labeled by ``$+$" and ``$-$". Then, the incidence matrix $\mathcal B$ for the ``labeled" graph is given as 
$\mathcal B _{i,k} = +1$ if prosumer $i$ is the positive end of the labeled transmission line $k \in  \mathcal E _ i$, $\mathcal B _{i,k} = -1$ if prosumer $i$ is the negative end of the labeled transmission line $k \in  \mathcal E _ i$, otherwise, $\mathcal B _{i,k} = 0$.
Suppose that prosumer $i$ is the positive end of the transmission line $k$, then, the dynamics of $I_{dk}$ and $I_{qk}$ are given by 
\begin{subequations}\label{line k}
\begin{align}
L_k \dot I _{dk} & = V_{di} - V_{dj} - R_k I _{dk} + \omega_0 L_k I_{qk} \\
L_k \dot I _{qk} & = V_{qi} - V_{qj}  - R_k I _{qk} - \omega_0 L_k I_{dk}.
\end{align}
\end{subequations}

\subsection{Behavior and Personal norm (BP) models}

\begin{figure}[t]
	\begin{center}
		\includegraphics[width= 0.32 \textwidth]{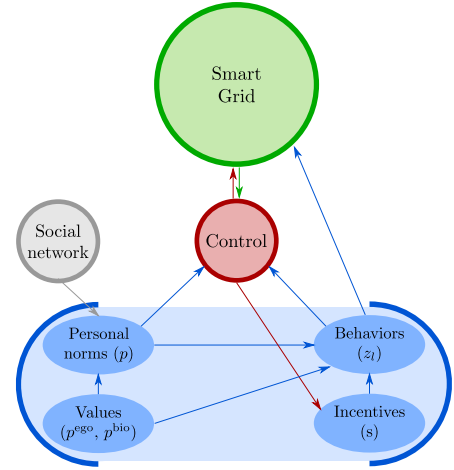}  \\
		\linespread{1}\caption{Schematic interactions within the proposed HPS framework.}   \label{interfig}
	\end{center}
\end{figure}

In this subsection, we focus on the modeling of prosumers' dynamics describing the consumption of energy. We will consider  \textbf{case i)} without social influence and \textbf{case ii)} with social influence. The interactions between the power grid, control scheme, behavior, personal norms, social influence, values and incentives are shown in Figure \ref{interfig}. To facilitate the presentation of our models, we first provide the definitions of personal values, egoistic and biospheric values, personal norms and financial incentives: 
	\begin{itemize}
		\item \emph{Personal values} reflect general and desirable life goals which are used as guiding principles to evaluate actions and situations on. Research has identified a set of universal values, meaning every individual endorses these values to some extend. However, individuals differ in how strongly they endorse each value. The more an individual endorses and prioritizes a value, the more influential this value is for someone's preferences and actions\cite{bouman2019motivating, bouman2018measuring, schwartz2012refining, steg2016values, stern1994value}. In case of energy saving behavior, two values appear of particular relevance: 
		\begin{itemize}
\item \emph{Egoistic values} concern goals to acquire possessions, money and status. 
Individuals with stronger egoistic values typically have weaker personal norms, which make them less likely to engage in pro-environmental actions such as energy saving behavior. Yet, energy saving behavior can also be associated with cost reductions, which may motivate individuals with stronger egoistic values to engage in energy saving behavior\cite{dogan2014making,schwartz2015advertising}.
\item \emph{Biospheric values} reflect goals to care about nature and the environment. Energy savings have clear environmental benefits, which is why stronger endorsement of biospheric values is typically indicative of stronger personal norms, and thereby stronger engagement in energy saving behavior\cite{dogan2014making}.
		\end{itemize}
		\item \emph{Personal norms} reflect a feeling of personal responsibility and feelings of moral obligation to take an action\cite{bouman2020insights}, e.g., 		
		the stronger an individual's personal norm to save energy, the more likely this individual is to engage in energy saving behavior (personal norms $\to$ behavior). 
			In the context of this paper, when someone strongly endorses biospheric and/or egoistic values, this individual is likely to experience a personal norm to take the corresponding actions (values $\to$ personal norms) \cite{stern2000new}. 
Then, for instance, individuals who strongly endorse biospheric values typically feel a stronger personal norm to take actions to save energy, eventually increasing the likelihood they will engage in energy saving behavior (values $\to$ personal norms $\to$  behavior).
	\item \emph{Financial incentives} include subsidies and financial rewards (e.g., lower energy bill) that can promote energy savings by making it more attractive. The impact of financial incentives is likely more pronounced for individuals with stronger egoistic values, as such individuals care relatively much about money and possessions\cite{dogan2014making,schwartz2015advertising}.
\end{itemize} 

\textbf{Case i)} In view of (\ref{physical i}), it is clear that the dynamic current loads depend on the dynamics of prosumers' behavior. 
Specifically, the BP dynamics of prosumer $i$ is represented by the following second order system 
\begin{subequations}\label{human i}
\begin{align}
\dot z_{li}   &= a_i( p_{i}  - z_{li}  -   h_i s_i )  \label{human zli} \\
 \dot p_i  & = c_i(   p^{\text{ego}}_i - p_i ) + d_i ( p^{\text{bio}}_i - p_i   ),   \label{human pi}
\end{align}
\end{subequations}
where $z_{li}$ and $p_i$ are the states for describing human activities (i.e., behavior and personal norms, respectively) and $s_i$ is the ``control input" representing incentives. The rationale behind this model becomes clear below. For a given $\bar s_i$, the steady-state solution of system \eqref{human i} satisfies
\begin{subequations}
\begin{align}
\bar z_{li} &= \bar p_i  - h_i \bar s_i  \label{steady zli} \\
\bar p_i & =  (c_i   p^{\text{ego}}_i  + d_i  p^{\text{bio}}_i  )  / (c_i + d_i). \label{steady pi}
\end{align}
\end{subequations}
It is clear that the compact form of (\ref{human i}) can be written as
\begin{subequations}\label{compact human}
	\begin{align}
	\dot z_l   &= A(p  - z_l   - H s ) \\
	\dot p   & = C(p^{\text{ego}}- p) + D (p^{\text{bio}} - p),
	\end{align}
\end{subequations}
where $A, H, C$ and $D$ are diagonal matrices, e.g. $A=\text{diag}(a_1, ..., a_N)$.
Detailed explanations about the variables and parameters of system (\ref{human i}) are given below.

\emph{Variables of prosumer $i$:}
\begin{itemize}
	
\item[a)] $z_{li}$: Degree of satisfaction of the load demand, with $0\le \bar z_{li}  \le 1$. It can be considered as the behavior of prosumer $i$ adjusting his/her own load $I_{Lgi} z_{li}$ ($g=d,q$). 
The closer $z_{li}$ is to 1, the more satisfied the load demand of prosumer $i$ is. This is a key variable that plays the role of interface between the grid (\ref{physical i}) and the prosumer's behavior (\ref{human i}). 

\item[b)] $p_i$: Personal norms, with $0 \le p_i   \le 1$. If we omit the ``control input" $-h_i   s_i$ in (\ref{human i}a) (see $h_i$ in item h) and $s_i$ in item c)), $p_i$ can be considered as the tracking reference for $z_{li}$, i.e., if $h_i   s_i = 0$, then $z_{li}  \to   p_i$ when time approaches infinity. This essentially describes the phenomenon for which the personal norms $p_i$ acts as a guide for the behavior $z_{li}$.

\item[c)] $s_i$: Financial incentives, with $ 0 \le   \bar s_i  \le \bar p_i /h_i$. $s_i$ represents the ``control input" (i.e., behavior interventions) to prosumer~$i$, influencing the behavior $z_{li}$ directly. Since $h_i  $ is semi-positive (see item h) in the following), $s_i \ge 0$ represents financial incentives that can motivate people to save energy. 
Some examples of such incentives can be subsidies and lower energy bills. 
For a given $\bar s_i$ and $\bar p_i$,  it follows from (\ref{steady zli}) that a larger value of $\bar s_i$ leads to a  smaller value of $\bar z_{li}$. For instance, this can describe the phenomenon for which higher subsidies generally lead to a larger energy saving of an individual, provided that his/her egoistic values are constant (see $h_i$ in item h)). From (\ref{human zli}) and (\ref{steady zli}), one can see that personal norms as well as incentives together influence one's energy saving behavior. Moreover, the inequality $\bar s_i  \le \bar p_i /h_i$ prevents that at the steady state the incentives are unreasonably high.  
\item[d)] $p^{\text{ego}}_i$: Egoistic values, with $0 \le p^{\text{ego}}_i  \le 1$. The influence of egoistic values on an individual is two-fold. First, egoistic values are negatively related to personal norms to save energy as pro-environmental behavior are often perceived as obstructing egoistic goals. The stronger the egoistic values are, the closer $p_i ^ {\text{ego}}$ to $1$ is, leading to a large value of $\bar p_i$ (see \eqref{steady pi}), which in turn implies a large value of $\bar z_{li}$ (see \eqref{steady zli}), i.e., more energy consumption. Second, egoistic values may also be positively related to energy saving behavior when energy savings are associated with financial benefits. Specifically, financial benefits (e.g., cost reductions) may promote energy saving behavior, and this effect will likely be larger for individuals with stronger egoistic values, see item h) for the modeling of its positive correlation to energy saving behavior.
\item[e)] $p^{\text{bio}}_i$: Biospheric values, with $0 \le p^{\text{bio}}_i  \le 1$. The stronger the biospheric values are, 
the closer $p^{\text{bio}}_i$ to $0$ is, leading to less energy consumption.

\end{itemize}

We assume $p^{\text{ego}}_i$ and $p^{\text{bio}}_i$ constant over the considered time windows because they change very slowly over time compared with the much faster dynamics of systems \eqref{physical i}--\eqref{human i}. 

\emph{Parameters of prosumer $i$:}
\begin{itemize}
\item[f)] $a_i > 0$ indicates the time constant of the behavior dynamics $z_{li}$. One can observe that a larger value of $a_i$ implies a faster response of $z_{li}$ to $p_i$ and $s_i$ variations. It is also clear that $a_i$ does not influence the steady state of $z_{li}$. 
\item[h)] $  0 \le h_i \le 1$, with $h_i=p^{\text{ego}}_i$, indicates the degree of influence of the incentives $s_i$ on the energy saving behavior $z_{li}$. 
Clearly, strong egoistic values ($p^{\text{ego}}_i \to 1$) imply larger values of $h_i$ ($h_i \to 1$). Moreover, one can observe from (\ref{steady zli}) that large values of $h_i$ indicate that incentives can have strong influence on motivating prosumer $i$ to save energy. Indeed, an individual with strong egoistic values does care about financial benefits, and hence is motivated by financial incentives to save energy. 
\item[i)] $c_i \ge 0 $ and $d_i \ge 0$ such that $c_i +  d_i > 0$ represent the weights of egoistic and biospheric values, respectively, indicating the degree to which each of these values influences the personal norms of prosumer $i$. 
From (\ref{steady pi}), one can observe that a relatively large value of $c_i$ implies $\bar p_i \to p^\text{ego}_i$, i.e., prosumer $i$ strongly endorses the egoistic values $p^\text{ego}_i$. On the other hand, a relatively large value of $d_i$ implies $\bar p_i \to p^\text{bio}_i$. 
\end{itemize}

\textbf{Case ii)}
Differently from case i), now we consider also the influence of social norms (i.e., perceptions that others save energy or expect you to save energy), which can motivate people to engage in sustainable energy-saving behavior. Then, the dynamics in (\ref{human pi}) become 
 \begin{align}
\dot p_i =   \sum_{ j \in \mathcal{S}_i} b_{ij }  (p_j - p_i)  + c_i( p^{\text{ego}}_i - p _i) + d_i(p^{\text{bio}}_i - p_i),
\end{align}
where $p_j $ with $j \in \mathcal{S}_i $ denotes the personal norms of the social neighbors of prosumer $i$, and $b_{ij} \in \mathbb R_{\ge 0}$ represents the weight of the social influence. It is clear that larger values of $b_{ij}$ imply that $\bar p_i$ is closer to $\bar p_j$.

Then, the compact model for the case with social influence is given by 
\begin{subequations}\label{compact human 1}
	\begin{align}
	\dot z_l   &=A ( p  - z_l   - H s ) \\
	\dot p   & = C(p^{\text{ego}}- p) + D (p^{\text{bio}} - p) - \mathcal L p, 
	\end{align}
\end{subequations} 
where $\mathcal L$ is the Laplacian matrix associated with the social network. Note that the social network topology is not necessarily identical to the physical topology of the power grid. One can observe that a relatively large $\mathcal L$ with respect to $C$ and $D$ corresponds to prosumers that are likely to achieve personal-norm consensus to some degree. In the following sections, we will mainly focus on case i). Then, we briefly extend the results to case ii).



According to some studies in environmental psychology \cite{dogan2014making,schwartz2015advertising}, our models include the possibility of promoting energy saving behavior in populations who more strongly endorse egoistic values by financial incentives ($s_i$ directly affects the behavior layer (\ref{human i}a)) in the short term ($a_i$ is generally larger than $b_{ij}$, $c_i$ and $d_i$). 
However, in the long term, incentives may be financially unsustainable and no longer effective when removed, implying that influencing individuals' personal norms (e.g. by strengthening biospheric values) may be more beneficial and effective.

\begin{remark}\label{opinion dynamics remark}
The proposed models are partially inspired by and consistent with the studies on opinion dynamics. First, they are partially consistent with some opinion dynamics models in which the ``topics" (state in models of one agent) are possibly more than one and logically correlated \cite{ye2020continuous}. 
Indeed, the proposed models have two correlated topics, i.e.,  $z_l$ and $p$, with the topic $p$ affecting the topic $z_l$. 
Second, the dynamics of topic $p$ are also partially consistent with the continuous-time Friedkin-Johnsen model\cite{friedkin1990social}, where an individual is stubborn about his/her opinion $p_i$ (depending on $p^{\text{ego}}$ and $p^{\text{bio}}$) and is also influenced by  others' opinions. \qedp
\end{remark}

\subsection{Human-physical system (HPS)}

In view of the dynamics of the physical AC grid in (\ref{physical i}) and (\ref{line k}), and prosumers' behavior and personal norms in (\ref{compact human}), the overall human-physical system (HPS) considering case i) is compactly written as 
\begin{subequations}\label{compact all}
\begin{align} 
C_t \dot V_d &= - R_L  ^ {-1} V_d  +  \omega_0 C_t  V_q + I_{td} + \mathcal B I_d - I _{Ld} z_l \\
C_t \dot V_q &= -  \omega_0 C_t  V_d  - R_L  ^ {-1} V_q  + I_{tq} + \mathcal B I_q - I _{Lq} z_l \\
L_t \dot I_{td} &= - V_d - R_t I_{td} + \omega_0 L_t I_{tq} + u_d  \\
L_t \dot I_{tq} &= - V_q - \omega_0 L_t I_{td} - R_t I_{tq}  + u_q  \\
L \dot I _d & = - \mathcal B ^ T V_d - R I _d + \omega_0 L I_q \\
L \dot I _q & = - \mathcal B ^ T V_q - \omega_0 L I_d - R I _q \\
  \dot z_l   &= A(p  - z_l   - H s)  \\
   \dot p   & = C(p^{\text{ego}}- p) + D (p^{\text{bio}} - p),
\end{align}
\end{subequations}
where vectors and matrices have appropriate dimensions. Specifically, (\ref{compact all}a)--(\ref{compact all}f) and (\ref{compact all}g), (\ref{compact all}h) represent physical and human dynamics, respectively, where the prosumers interact with the power grid by changing their energy saving behavior ($z_l$). Moreover, $u_d$ and $u_q$ represent the inputs to control the power grid, while $s$ represents the ``control input" (i.e., behavior interventions) to influence the behavior of the prosumers. Similarly, one can obtain the corresponding HPS considering case ii) by replacing (\ref{compact all}g) and (\ref{compact all}h) with (\ref{compact human 1}a) and (\ref{compact human 1}b), respectively.
In the following sections, we will mainly focus on the HPS in (\ref{compact all}), i.e., case i). Then, we briefly extend the results to case ii).


\section{Optimization problem}

In this section we present the control objectives of this work and formulate a social-physical welfare optimization problem.

\subsection{Physical welfare}\label{subsec:physical_w}
It is convenient in practice to decouple
the control of the active power $P_{i} = \frac{3}{2} (V_{di} I_{tdi}+ V_{qi} I_{tqi}) $ from the control of the reactive power $Q_{i} = \frac{3}{2} (V_{qi} I_{tdi} - V_{di} I_{tqi}) $. One possible way to do this is to regulate the $q$-component of the voltage to zero. Specifically, it is desirable to achieve
\begin{align}\label{steady Vq}
\lim_{t \to \infty} V_{qi}(t) = 0, \,\,\,\, i = 1, 2, ..., N. 
\end{align}
It is then evident that the active power $P_{i}$ and reactive power $Q_{i}$ depend on the $d$-component of the voltage $V_{di}$, and the currents $I_{tdi}$ and $I_{tqi}$, respectively.

Given any constant $\bar u_d$, $\bar u_q$ and $\bar z_l$, and considering (\ref{steady Vq}), the steady-state of the power grid  (\ref{compact all}a)--(\ref{compact all}f) satisfies
\begin{subequations}\label{steady physical ac}
	\begin{align}
	\bar V _d = & - R _ t \bar I _ {td} + \omega_0 L_ t  \bar I _{tq} + \bar u _ d \\ 
	\mathbf 0   = & -\omega_0 L _ t \bar I _ {td} -  R _ t \bar I _ {tq} + \bar u _ q \\
	\bar I _{td}= & ~I _ {Ld} \bar z_l - (- R_L ^ {-1} + \mathcal B J ) \bar V_d   \\
	\bar I _ {tq} = & ~I _ {Lq} \bar z_l -  (- \omega_0 C_t  + \mathcal B K  ) \bar V_d,   
	\end{align}
\end{subequations}
where we use $\bar I _d = (-R - \omega_0 ^ 2 LR^{-1}L)^{-1} \mathcal B ^T \bar V_d = J \bar V_d$ and $\bar I _ q = - \omega_0 R ^ {-1} L \bar I _ d = K \bar V_d $, with $ J :=  (-R - \omega_0 ^ 2 LR^{-1}L)^{-1} \mathcal B ^T $ and $ K := -\omega_0 R ^ {-1} L J $. 
We also notice that the steady-state of the power grid depends on $\bar z_l$. Then, given any constant input $\bar s$, the steady-state of the of human behavior (\ref{compact all}g), (\ref{compact all}h) satisfies
\begin{subequations}\label{steady human ac}
	\begin{align}
	\bar z_l  &= \bar p  - H \bar s \\ 
	\bar p  &=  (C+D ) ^ {-1} (C p^{\text{ego}} + D p^{\text{bio}} ).
	\end{align}
\end{subequations}
From (\ref{steady physical ac}) and (\ref{steady human ac}), one can verify that given $(\bar u_d , \bar u_q , \bar s)$, the forced equilibrium of the HPS (\ref{compact all}) exists and is unique. 
Also, in view of $\bar I_{td}$ in (\ref{steady physical ac}c), one has 
\begin{align}\label{current matching}
\sum_{i=1}^{N} (I_{Ldi} \bar z_{li} + \bar V_{di}/R_{Li}) = \mathbf 1 ^ T \bar I_{td}, 
\end{align}
which implies that the total (active) current load is equal to the total (active) generated current, leading to the formulation of the current sharing objective. Indeed, to avoid the overstressing of one or more energy sources, it is desirable in practice that the overall current generation is shared among all the prosumers  proportionally to their generation capacities. This desire is equivalent to achieving $\pi_{ci}  \bar I _{tdi}  = \pi _{cj} \bar I _{tdj},  i,j \in \mathcal N$, where $\pi_{ci} \in \mathbb R _{>0}$ and $\pi_{cj} \in \mathbb R _{>0} $ are constant weights depending on the generation capacity of  prosumers $i$ and $j$, respectively. For example, a relatively large $\pi_{ci}$ corresponds to a relatively small generation capacity.
Therefore, to achieve this goal, in analogy with \cite{cucuzzella2019distributed}, we assign to every
prosumer a strictly convex quadratic \lq cost' function depending on the generated current $I_{tdi}$. Then, the overall cost can be expressed as 
\begin{align}
C(I_{td})=  \sum \frac{1}{2} \pi _{ci} {I _{tdi}  } ^2.  
\end{align}
Note that the current sharing objective may also be  interpreted as an action to pursue ``fairness" in sharing tasks among all the prosumers, which can potentially enhance the energy efficiency by cooperating with each other.


Now, noticing that (\ref{steady Vq}) implies that the voltage magnitude $|\bar V_i | := (\bar V_{di}^2 + \bar V_{qi} ^ 2) ^ {1/2}$ is determined only by $V_{di}$, we consider the regulation of $V_d$. Let $V_r= [V_{r1}\,\,\cdots\,\, V_{rN}]^T \in \mathbb R ^{N}_{>0}$ be the vector of  voltage references for $V_d$.  
Ideally, we would like to achieve exact voltage regulation, i.e.,
\begin{align}\label{steady Vd}
\lim_{t \to \infty} V_{di}(t) = \bar V_{di} = V_{ri}, \,\, i = 1, 2, ..., N.
\end{align}
However, recalling that in low-voltage grids there is a strong coupling between the active power flows 
and the voltage magnitude \cite{guerrero2007decentralized,golsorkhi2016gps}, in order to achieve current sharing, deviations from the  voltage references are unavoidable.
Then, to regulate the voltages sufficiently close to the corresponding references, we  aim at minimizing the voltage error, considering the strictly convex quadratic \lq cost' function
$\| V_d- V_r\| ^ 2$.

%
%
%
%
%
%

\subsection{Social welfare}\label{subsec:social_w}
In this subsection, we formulate the social (and economic) welfare problem within the considered power grid.  

First, let $I_{Li} := (I_{Ldi} ^ 2 + I_{Lqi} ^ 2) ^{1/2}$. Then, we assign to every
prosumer a strictly concave
quadratic \lq utility' function $U_i( z_{li})$ depending on the energy saving behavior of prosumer $i$. Then, the overall utility can be expressed as
\begin{align}\label{U z_l}
U (z_l ) = - \sum _{i \in \mathcal{N}} \frac{1}{2} \pi _{ui} (I  _{Li} (1- z_{li}) )^ 2,
\end{align} 
where the parameter $ \pi _{ui} \in  \mathbb R _{\ge 0}$ weights the satisfaction of the load demand of prosumer $i$.
For example, a relatively large $\pi _{ui}$ corresponds to a relatively large request of comfort from prosumer $i$. By minimizing $-U(z_l)$, we aim at satisfying prosumers' load demands as much as possible by making $\bar z_l$ close to $ \mathbf 1 $ (i.e., making $1 - \bar z_{li}$ close to $0$).

%

We also would like to minimize the amount of incentives $s$, which we assume to be proportional to the square of the incentives themselves. Thus, we aim at minimizing the cost function $\|s\|^2$.

%
%
%
%
%
%
%


\subsection{Social-physical welfare}
Considering the physical and social welfares introduced in Subsections \ref{subsec:physical_w} and \ref{subsec:social_w}, respectively, we now formulate the overall social-physical welfare optimization problem. Let the optimization variables be denoted by the superscript $^*$ and let $\hat x_c := [{ z_l^ * }^T \, {I_{td} ^ *}^T \, {I_{tq} ^ *}^T \,  {u_d^*}^T  \, {u_q ^ *}^T \, {V_d ^ *}^T \, 
{ s^*}^T]^T \in \mathbb R^{7N} $ be the vector of the optimization variables.
Then, consider the following convex minimization problem:
\begin{subequations}\label{opti prob}
	\begin{align}
	\underset{\hat x_c}{\min} & \quad \mathcal F(z_l^ * , I_{td} ^ * , u_d^*, u_q ^ * ,  V_d ^ * , s^*)     \\
	\text{s.t. }  & \quad \mathbf 0= I _ {Ld} \bar z_l ^* - \bar I _{td}^* - (- R_L ^ {-1} + \mathcal B J ) \bar V_d^*   \\
	&  \quad  \mathbf 0 = I _ {Lq}\bar z_l  ^*  -  \bar I _ {tq} ^*  -  (- \omega_0 C_t  + \mathcal B K  ) \bar V_d ^* \\
	& \quad \mathbf 0=\bar V _d  ^* + R _ t \bar I _ {td} ^*  -\omega_0 L_ t  \bar I _{tq} ^* - \bar u _ d ^* \\ 
	& \quad  \mathbf 0=-\omega_0 L _ t \bar I _ {td}  ^* -  R _ t \bar I _ {tq} ^* + \bar u _ q ^*    \\
	& \quad \mathbf 0=\bar p  - \bar z_l  ^*  - H \bar s  ^*, 
	\end{align} 
\end{subequations}
where $\bar p$ is given by (\ref{steady human ac}b), and
\begin{align}\label{eq:obj_function}
\mathcal F := &~\frac{1}{2}  \alpha \sum \pi _{ui} I_{Li} ^ 2(1-z_{li} ^*)^2 + \frac{1}{2} \beta  \sum \pi_{ci} (I _{tdi} ^ * ) ^ 2   \nonumber \\
& + \frac{\gamma}{2}  \|u_d ^* \| ^ 2 + \frac{\delta}{2}  \|u_q ^* \| ^ 2 + \frac{ \epsilon }{2}\|V_d ^*  - V_r \|^ 2+ \frac{\eta }{2} \|s ^* \| ^ 2,   
\end{align}
where $\alpha, \beta, \gamma$, $\delta$, $\epsilon$ and $\eta$ are positive constants that can be chosen to prioritize one objective over another. 
We also note that the equality constraint (\ref{opti prob}e) implies $\bar V_q ^* = \mathbf 0$, satisfying~\eqref{steady Vq}, and in \eqref{eq:obj_function} the terms depending on $u_d^*$ and $u_q^*$ concern the minimization of the control efforts. Furthermore, we note that in addition to the equality constraints (\ref{opti prob}b)--(\ref{opti prob}f), inequality constraints (see e.g. \cite{stegink2016unifying}) may also be considered for instance to guarantee that voltages converge within a band allowing for a safe and proper functioning of the prosumers' appliances or to avoid too high incentives that are financially unsustainable. However, for the sake of exposition and due to the page limitation, we will not include inequality constraints in the following analysis.

\section{Controller and stability analysis}
In this section, we will present the design of a primal-dual controller to solve the optimization problem (\ref{opti prob}).

\subsection{Design of a primal-dual controller }
Let $\lambda:=[\lambda _ a^T \, \lambda _b^T \, \lambda_c^T \,  \lambda_d^T \,  \lambda_e^T ]^T \in \mathbb R^{5N}$ denote the vector of the Lagrange multipliers corresponding to the constraints in (\ref{opti prob}b)--(\ref{opti prob}f). 
Moreover, let $x_c := [\hat x_c^T \,\lambda^T]^T \in \mathbb R^{12N}$ be the state vector of the primal-dual controller we will design later in this subsection. Then, the Lagrangian function corresponding to the optimization problem (\ref{opti prob}) is the following
\begin{align}
\ell (x_c):=&~\mathcal F(z_l^ * , I_{td} ^ * , u_d^*, u_q ^ * ,  V_d ^ * , s^*) \nonumber \\
&+ \lambda_a ^ T (I _ {Ld}   z_l  ^* -   I _{td} ^*  - (-R_L ^ {-1}  + \mathcal B J  )  V_d ^*) \nonumber\\
& +\lambda_b ^ T ( I _ {Lq}   z_l ^* -   I _ {tq} ^*  +  (  \omega_0 C_t   - \mathcal B K  )   V_d ^*  ) \nonumber\\
& + \lambda_c ^ T (  V _d ^*  + R _ t   I _ {td}^*  - \omega_0 L_ t    I _{tq} ^*  -    u _ d ^* ) \nonumber\\
& + \lambda_d ^ T ( -\omega_0 L _ t   I _ {td} ^*  - R _ t  I _ {tq} ^*  +   u _ q ^*    ) \nonumber\\
& + \lambda_e ^ T (   \bar p - z_l   ^* - H   s ^* ).
\end{align}
Thus, the first order optimality conditions for the optimization problem (\ref{opti prob}) are given by the Karush-Kuhn-Tucker (KKT) conditions, i.e.,
\begin{subequations}\label{KKT}
	\begin{align}
	& \mathbf 0 = - \alpha \Pi_ u I_L ^ 2 (\mathbf 1- \bar z_l ^ *  ) + I_{Ld} \bar \lambda_a + I _{Lq} \bar  \lambda_b + \bar \lambda_e\\
	& \mathbf 0 = \beta \Pi_c \bar I_{td} ^* - \bar \lambda_a + R_t \bar  \lambda_c - \omega_0 L_t \bar \lambda_d  \\
	& \mathbf 0 =   - \bar \lambda_b - \omega_0 L_t \bar \lambda_c - R_t \bar \lambda_d  \\
	& \mathbf 0 = \gamma \bar u_d ^*  - \bar  \lambda_c    \\
	& \mathbf 0 = \delta \bar u_q ^* + \bar \lambda_d    \\
	& \mathbf 0 = \epsilon (\bar V_d ^* - V_r) - (-R_L ^ {-1} + \mathcal B J ) ^ T  \bar \lambda_a   \nonumber \\
	& \quad \,\,\,  + (\omega_0 C_t - \mathcal BK) ^ T \bar \lambda_b + \bar \lambda_c  \\
	& \mathbf 0 = \eta \bar s ^*  - H \bar \lambda_e     \\
	& \mathbf 0 = I_{Ld} \bar z_l ^* - \bar I _{td}^*  - (-R_L ^ {-1}+ \mathcal B J ) \bar V_d ^*   \\
	& \mathbf 0 = I _{Lq} \bar z_l ^* - \bar I_{tq}^*  + (\omega_0 C_t - \mathcal B K ) \bar V_d^*  \\
	& \mathbf 0 = \bar V_d ^* + R _t \bar I _{td} ^* - \omega_0 L_t \bar I_{tq} ^* - \bar u_d^*  \\
	& \mathbf 0 = -\omega_0 L_t \bar I _{td} ^* - R_t \bar I_{tq} ^*  + \bar u_q ^*  \\
	& \mathbf 0 =  \bar p   - \bar z_l ^* - H \bar s ^*,
	\end{align}
\end{subequations}
with $\Pi_u = \text{diag}(\pi_{u1}, \cdots, \pi_{uN})$, $\Pi_c = \text{diag}(\pi_{c1}, \cdots, \pi_{cN}) $ and $I_L ^ 2= \text{diag}(I_{L1}^2, \cdots, I_{LN}^2)$.
Since the optimization problem (\ref{opti prob}) is convex, then strong duality holds \cite{boyd2004convex}. Thus, $\bar z_l^ * , \bar I_{td} ^ * , \bar I_{tq} ^ *, \bar u_d^*, \bar u_q ^ * ,  \bar V_d ^ * , \bar s^*$ are optimal if and only if there exist $\bar \lambda _ a, \bar \lambda _b, \bar \lambda_c, \bar \lambda_d$ and $\bar \lambda_e$ that satisfy (\ref{KKT}).

Based on the KKT conditions in (\ref{KKT}) and under the assumption that each controller can exchange
information among its neighbors through a communication
network with the same topology as the physical
network, we design the following distributed control scheme by using the primal-dual dynamics of the optimization problem (\ref{opti prob}), i.e., 
\begin{subequations}\label{controller}
	\begin{align}
	-\tau_z \dot z_l ^ * &= - \alpha \Pi_ u I_L ^ 2 (1- z_l ^ *  ) + I_{Ld} \lambda_a + I _{Lq} \lambda_b + \lambda_e  \\
	-\tau_{td} \dot I _ {td}  ^ * &= \beta \Pi_c I_{td} ^ *  - \lambda_a + R_t \lambda_c - \omega_0 L_t \lambda_d \\
	-\tau_{tq} \dot I _ {tq} ^ *  &=  - \lambda_b - \omega_0 L_t \lambda_c - R_t \lambda_d \\
	-\tau_{ud} \dot u _ d ^ * &= \gamma u_d ^ *  - \lambda_c + p_A \\
	-\tau_{uq} \dot u _ q ^ *  &= \delta u_q ^ * + \lambda_d + p_B   \\
	-\tau_{V} \dot V _ d ^ *  &= \epsilon (V_d ^ * - V_r) - (-R_L ^ {-1} + \mathcal B J ) ^ T \lambda_a  \nonumber \\
	&\,\,\,\,+ (\omega_0 C_t - \mathcal BK) ^ T \lambda_b + \lambda_c  \\
	-\tau_{s} \dot s ^ * & = \eta s ^ * - H \lambda_e + p_C   \\
	\tau_a \dot \lambda_a & = I_{Ld} z_l ^ * - I _{td} ^* - (-R_L ^ {-1}+ \mathcal B J ) V_d  ^*  \\
	\tau_b \dot \lambda_b & = I _{Lq} z_l ^ * - I_{tq}^* + (\omega_0 C_t - \mathcal B K ) V_d ^* \\
	\tau_c \dot \lambda_c & = V_d ^ *  + R _t I _{td} ^ * - \omega_0 L_t I_{tq} ^ * - u_d ^ *  \\
	\tau_d \dot \lambda_d & = -\omega_0 L_t I _{td} ^ * - R_t I_{tq} ^ * + u_q ^ *   \\
	\tau_e \dot \lambda_e & = \bar  p    - z_l  ^ * - H s^ *, 
	\end{align}
\end{subequations}
where $\tau _z, \tau_{td}, \tau_{tq}, \tau_{ud}, \tau_{uq}, \tau _{V}, \tau _s, \tau_a, \tau_b , \tau_c, \tau_d, \tau_e \in \mathbb R ^{N \times N}$ are positive diagonal matrices, which can be tuned to adjust the controller response. Moreover, the vectors $p_A$, $p_B$ and $p_C$ are the controller input ports, which will be used in the next subsection to interconnect the controller (\ref{controller}) with the HPS (\ref{compact all}). 
We also note that since in the optimization problem (\ref{opti prob}) the objective function \eqref{eq:obj_function} is quadratic with respect to $z_l^ * , I_{td} ^ * , u_d^*, u_q ^ * ,  V_d ^ *, s^*$ and the constraints (\ref{opti prob}b)--(\ref{opti prob}f) are linear, then, given constant $\bar p_A,  \bar p_B$ and $\bar p_C$, it can be shown that the solution $\bar x_c$ to (\ref{controller}) is unique.

\subsection{Stability analysis}

In this subsection, we will show that the HPS (\ref{compact all}) in closed loop with the primal-dual controller \eqref{controller} is stable and converges to the solution of the optimization problem (\ref{opti prob}).

Let $\tilde p  :=  \bar p - p = (C+D)^{-1} (C p^{\text{ego}} + D p^{\text{bio}} ) - p$, with  $\bar p$ given by (\ref{steady human ac}b). Then, the human dynamics (\ref{compact all}g), (\ref{compact all}h) can be rewritten as 
\begin{subequations}\label{trans human}
	\begin{align}
	\dot z_l &= -A z_l + A( \bar p - \tilde p  - H s     )  \\
	\dot {\tilde p}   &= - (C+D) \tilde p. 
	\end{align}
\end{subequations}
It is clear that $-A$ and $-(C+D) $ are Hurwitz. Thus, given any two diagonal and positive definite matrices $Q_1$ and $Q_2$, let $P_1$ and $P_2$ denote the corresponding unique solutions of the Lyapunov equations
\begin{align}
& -A^T P_1 - P_1 A + Q_1 = \mathbf 0 \label{P1}\\
& -  (C+D)^T P_2 - P_2 (C+D) + Q_2 = \mathbf 0, \label{Lya equ}
\end{align}
where $P_1$ and $P_2$ are positive definite matrices.
Then, we interconnect the controller (\ref{controller}) with the HPS (\ref{compact all}) by choosing 
\begin{align}\label{port0}
u_d = u_d ^* , \,\, u_q = u_q ^ * , \,\, s= s^*
\end{align}
and
\begin{align}\label{port}
p_A= I_{td}, \,\, p_B= I_{tq},\,\, p_C= -2H^T A^T P_1 ^ T z_l. 
\end{align}

Let $x_s := [V_d ^T \,\, V_q^T\,\, I_{td}^T\,\,I_{tq}^T\,\, I_d^T\,\,I_q^T \,\,z_l^T\,\, \tilde p^T]^T \in \mathbb R^{6N+2E}$ denote the state of the HPS consisting of (\ref{compact all}a)--(\ref{compact all}f) and (\ref{trans human}). 
Now we are ready to present the main result of this paper.

\begin{theorem}\label{Theorem 1}
	The closed-loop system (\ref{compact all}), (\ref{controller}), (\ref{port0}) and  (\ref{port}) converges to an equilibrium solving \eqref{opti prob}.  
\end{theorem}

\emph{Proof.}
We take three steps to conduct the proof. 

\emph{Step 1.}
In this step, we first propose the following storage function \cite{kosaraju2020differentiation} for the HPS (\ref{compact all}a)--(\ref{compact all}f) 
\begin{align}
S_{p} = &\,\, \frac{1}{2}( \dot V_d ^  T C _ t\dot V_d +   \dot V_q ^  T C _ t  \dot V_q +  \dot I_{td} ^ T  L_t  \dot I_{td} +     \dot I_{tq} ^ T   L_t   \dot I_{tq}  \nonumber\\
& +  \dot I_d ^T  L_t   \dot I_d +  \dot I_q ^T  L_t    \dot I_q ),
\end{align}
which satisfies 
\begin{align}
\dot S_p = & -   \dot V_d ^T  R_L^{-1}   \dot V_d  -   \dot V_q ^T  R_L ^ {-1}  \dot V_q 
-  \dot I_{td} ^ T  R_t  \dot I_{td}  -     \dot I_{tq} ^ T   R_t   \dot I_{tq}  \nonumber\\
& -  \dot I_d ^T  R   \dot I_d  -  \dot I_q ^T  R    \dot I_q 
-  \dot V_d ^ T I _{Ld} \dot z_l   -   \dot V_q ^ T I _{Lq} \dot z_l  \nonumber\\
& + \dot I _{td}^T \dot u_d ^ * + \dot I_{tq}^T \dot u_q ^*
\end{align}
along the solutions to (\ref{compact all}a)--(\ref{compact all}f). 
Now, supposing without loss of generality that the loads absorb positive reactive power (i.e., the loads are predominantly inductive rather than capacitive), $I_{Lq}$ is negative definite. Then, by virtue of the Young's inequality \cite{hardy1952inequalities}, we have
\begin{align}
-  \dot V_d ^ T I _{Ld} \dot z_l   -   \dot V_q ^ T I _{Lq} \dot z_l  
\le&~ \dot V_d  ^ T \frac{I_{Ld}}{2 \zeta_2} \dot V_d  + \dot z_l ^ T  \frac{\zeta_2 I_{Ld} }{2} \dot z_l  \nonumber\\
& - \dot V_q ^T \frac{I_{Lq}}{2 \zeta_3} V_q ^ T - \dot z_l ^ T \frac{\zeta_3 I_{Lq}}{2} \dot z_l,
\end{align}
with $\zeta_2$ and $\zeta_3$ being arbitrary positive reals. 

Now, for the human dynamics in (\ref{trans human}) we propose the following storage function
\begin{align}
S_h = \dot z_l ^ T P_1 \dot z_l +\dot  {\tilde p }^ T P_2  \dot  {\tilde p },
\end{align}
which satisfies
\begin{align}
\dot S_h = &- \dot z_l ^T Q_1  \dot z_l    
- \dot {\tilde p} ^ T Q_2  \dot {\tilde p}   
- 2 \dot z_l ^ T P_1 A ( \dot {\tilde p} + H \dot s^*),   
\end{align}
along the solutions to (\ref{trans human}). 
By virtue again of the Young's inequality, one has
\begin{align}
- 2 \dot z_l^T P_1 A ( \dot {\tilde p} + H \dot s ^*)  
\le &~\dot z_l ^T \frac{P_1 A}{\zeta_1} \dot z_l  + \dot {\tilde p } ^T  \zeta_1 P_1 A \dot {\tilde p}\nonumber \\
   &- 2 \dot z_l ^ T P_1 A   H \dot s^*, 
\end{align}
with $\zeta_1$ being an arbitrary real. 
Then, for the overall HPS (\ref{compact all}a)--(\ref{compact all}f) and (\ref{trans human}), the storage function $S_{ph}:=S_p+S_h$ satisfies
\begin{align}\label{dot HPS}
\dot S_{ph} \le&  - \dot V_d  ^ T \left(R _ L ^{-1} - \frac{1}{2 \zeta_2} I_{Ld}\right)\dot V_d   \nonumber\\
&  - \dot V_q  ^ T \left(R _ L ^{-1} + \frac{1}{2 \zeta_3} I_{Lq}\right)V_q  \nonumber\\
&  - \dot z_l ^T \left(Q_1 - \frac{1}{\zeta_1}P_1 A -  \frac{\zeta_2}{2}I_{Ld} + \frac{\zeta_3}{2}I_{Lq}\right) \dot z_l \nonumber\\
&  - \dot {\tilde p} ^ T  \left(Q_2  - \zeta_1 P_1 A\right) \dot {\tilde p}  \nonumber\\
& +  \dot I_{td} ^ T \dot u_d ^* + \dot I_{tq} ^ T \dot u_q ^* - 2 \dot z_l ^ T P_1 A   H \dot s ^*,  
\end{align}
along the solution to (\ref{compact all}a)--(\ref{compact all}f) and (\ref{trans human}). We observe that the terms in the first and second lines can be made nonpositive by selecting sufficiently large $\zeta_2$ and $\zeta_3$, respectively. With the selected $\zeta_2$ and $\zeta_3$, also the terms in the third line can be made nonpositive by selecting sufficiently large $Q_1$ and $\zeta_1$. With the selected $\zeta_1$, it is finally possible to make also the terms in the fourth line nonpositive by selecting a sufficiently large $Q_2$. Then, we have
\begin{align}\label{passivity HPS}
\dot S_{ph} \le    \dot I_{td} ^ T \dot u_d  ^*+ \dot I_{tq} ^ T \dot u_q  ^* - 2 \dot z_l ^ T P_1 A   H \dot s ^*, 
\end{align}
which implies the HPS (\ref{compact all}a)--(\ref{compact all}f) and (\ref{trans human}) is passive with respect to the supply rate $[\dot I_{td}^T\,\, \dot I_{tq}^T\,\, -2 \dot z_l  ^ T P_1AH ] [\dot u_d^{*T}\,\, \dot u_q^{*T}\,\, \dot s^{*T}]^T$ and storage function $S_{ph}$.

\emph{Step 2. }
In this step, we propose for the primal-dual controller (\ref{controller}) the following storage function
\begin{align}
S_c =&~\frac{1}{2}(\dot z_l ^ {*T} \tau_z \dot z^ *  _l
+ \dot I_{td} ^{*T} \tau_{I_{td}} \dot I_{td} ^ *   +   \dot I_{tq} ^{*T} \tau_{I_{tq}} \dot I_{tq} ^ * 
+ \dot u_d ^ {* T} \tau _{u_d} \dot u _d ^ *   \nonumber\\
&+\dot u_q ^ {* T} \tau _{u_q} \dot u _q ^ *   +    \dot V_d ^ {*T} \tau_{V} \dot V_d ^ *  
+ \dot s ^ {* T }  \tau _ s \dot s ^ *   + \dot \lambda_a ^ T \tau_a \dot \lambda_a \nonumber\\
&+\dot \lambda_b ^ T \tau_b \dot \lambda_b + \dot \lambda_c  ^ T \tau_c \dot \lambda_c 
+ \dot \lambda_d ^ T \tau_d \dot \lambda_d + \dot \lambda_e ^ T \tau_e \dot \lambda_e ),
\end{align}
which satisfies
\begin{align}\label{dot controller}
\dot S_c = \, &    - \alpha\dot z_l ^ {*T}  \Pi_u I_L ^ 2  \dot z^ *  _l 
- \beta\dot I_{td} ^{*T}  \Pi_c  \dot I_{td} ^ *   
- \gamma\dot u_d ^ {* T}  \dot u _d ^ *   \nonumber\\
& -  \delta\dot u_q ^ {* T}   \dot u _q ^ *   -   \epsilon \dot V_d ^ {*T}  \dot V_d ^ *  
- \eta\dot s ^ {* T }   \dot s ^ *   \nonumber\\
& - \dot u_d ^{* T}  \dot p_A - \dot u_q ^ {*T} \dot p_B - \dot s ^{*T} \dot p_C  \nonumber\\
\, \le & - \dot u_d ^{* T}  \dot p_A  - \dot u_q ^ {*T} \dot p_B - \dot s ^{*T} \dot p_C 
\end{align}
along the solutions to (\ref{controller}), implying that the controller (\ref{controller}) is passive with respect to the supply rate $-[\dot p_{A}^T\,\, \dot p_{B}^T\,\, \dot p_C^T][\dot u_d^{*T}\,\, \dot u_q^{*T}\,\, \dot s^{*T}]^T$ and storage function $S_c$, where $p_A$, $p_B$ and $p_C$ are given in (\ref{port}). 

\emph{Step 3.}
As the last step,  for the closed-loop system  (\ref{compact all}a)--(\ref{compact all}f), (\ref{controller}), (\ref{trans human}) with (\ref{port0}) and  (\ref{port}), we propose $S:=S_{ph}+S_c$ as storage function, which satisfies $\dot S \le 0$ along the solutions to (\ref{compact all}a)--(\ref{compact all}f), (\ref{controller}), (\ref{trans human}) with (\ref{port0}) and  (\ref{port}), implying that such solutions are bounded.
Therefore, there exists a forward invariant set $\Omega$ and by LaSalle's invariance principle the solutions that start in $\Omega$ converge to the largest invariant set contained in
\begin{align}
\begin{split}
\Omega \cap \{(x_s,{x}_c) \in \mathbb R^{18N+2E}|&~\dot{x}_s={\bf0}, \dot{z}_l^*={\bf0}, \dot{I}_{td}^*={\bf0}, \dot{u}_d^*={\bf0},\\
&~\dot{u}_q^\ast={\bf 0}, \dot{V}_d^\ast={\bf0},\dot{s}^\ast={\bf0}\}.
\end{split} \nonumber
\end{align}
Then, from  (\ref{controller}d), (\ref{controller}j), (\ref{controller}e), (\ref{controller}c), (\ref{controller}b) and (\ref{controller}a) it follows that on the largest invariant set $\dot{\lambda}_c, \dot{I}_{tq}^*, \dot{\lambda}_d, \dot{\lambda}_b, \dot{\lambda}_a$ and $\dot{\lambda}_e$ are also equal to zero, respectively. Moreover, we observe  from (\ref{trans human}b) that on the largest invariant set $p=\bar p$,  with  $\bar p$ given by (\ref{steady human ac}b).
Finally, observing from \eqref{steady physical ac} and \eqref{steady human ac} that $\bar x_s$ is uniquely determined by $\bar u_d = \bar{u}_d^\ast$, $\bar u_q = \bar{u}_q^\ast$ and $\bar s = \bar{s}^\ast$, we can conclude from (\ref{controller}h)--(\ref{controller}l) that at the steady-state the physical state variables coincide with the corresponding optimization variables. 
\qedp


\subsection{HPS considering social influence}
In the last two subsections, we designed the primal-dual controller and analyzed the state convergence of the closed-loop systems for case i), which does not consider social influence. 
In this subsection, we briefly extend our results to case ii), which includes social influence on people's activities (in (\ref{compact human 1})). Thus, the resulting HPS model consists of (\ref{compact all}a)--(\ref{compact all}f) and (\ref{compact human 1}). 

In view of the model in (\ref{compact human 1}), one can verify that its steady state satisfies 
\begin{align}\label{compact human 2 steady}
\bar p &= (C+D+\mathcal L) ^ {-1} (C p^{\text{ego}} + D p^{\text{bio}}) \\
\bar z_l &= \bar p - H \bar s.
\end{align}
The primal dual controller in (\ref{controller}) is still applicable to the new HPS. For the stability analysis, some changes are needed and briefly explained in the following. Let $\tilde p = \bar p - p =  (C+D+\mathcal L)^{-1} (C p^{\text{ego}} + D p^{\text{bio}} ) - p$, then the transformed system can be written as 
\begin{subequations}
	\begin{align}
	\dot z_l &= -A z_l + A( (C+D + \mathcal L)^{-1}  (C p^{\text{ego}} + D p^{\text{bio}}  )- \tilde p  - H s     )  \nonumber\\
	\dot {\tilde p}   &= - (C+D + \mathcal L) \tilde p \nonumber. 
	\end{align}
\end{subequations}
One needs to replace $C+D$ by $C+D+ \mathcal L$ in the Lyapunov equation in (\ref{Lya equ}), where $C+D+ \mathcal L$ is still a Hurwitz matrix. Then, the rest of the analysis can be conducted analogously to the proof of Theorem 1. Therefore, the following corollary holds.

\begin{corollary}
	The closed-loop system (\ref{compact human 1}), (\ref{compact all}a)-(\ref{compact all}f), (\ref{controller}), (\ref{port0}) and  (\ref{port}) converges to an equilibrium solving \eqref{opti prob}.   \qedp
\end{corollary}

 \begin{table}[t]
	\centering
	\caption{Coefficients of prosumers and lines}
	\begin{tabular}{  p{1.5cm}||p{1cm}|p{1cm}|p{1cm} | p{1cm} }
		& Pros. 1 & Pros. 2 & Pros. 3   &  Pros. 4 \\
		\hline
		$C_{t}$ ($\mu$F)  & 62.86 &62.86 &62.86 &62.86\\
		$L_{t}$  (mH) &  2.1 & 2.0 & 1.9 &1.8 \\
		$R_{t}$ (m$\Omega$)   &40.2 &38.7 &34.6 &31.8 \\
		$R_L$  ($\Omega$) 			&16.9 &13 &10.9 &13\\
		$I_{Ld}$   (A) & 30 &  25 &  30 & 26 \\
		$I_{Lq}$    (A) & -20 &  -15 &  -10 & -18\\		
		$V_r$ (V) & 120 $\sqrt{2}$  & 120 $\sqrt{2}$ & 120 $\sqrt{2}$ & 120 $\sqrt{2}$\\
		$\pi_{ci}$  & 1 & 1 &1 &1 \\
		$\pi _{ui}$ & 1&1 &1&1\\
		\hline
		\hline
		& Line 1 & Line 2 &  Line 3 &  Line 4 \\
		\hline
		$R_{k}$ ($\Omega$)   &  0.25  & 0.27  & 0.24   &  0.26 \\
		$L_{k}$   ($\mu$H) & 	1.2  &  1.3  & 1.8   & 2.1 \\
		\hline
	\end{tabular}
	\label{table 2}
\end{table}

\section{Simulations}
We present simulation results in this section. Specifically, we consider an AC microgrid consisting of four prosumers. Both the human models in (\ref{compact human}) (case i)) and (\ref{compact human 1}) (case ii)) are considered. 

The parameters of the AC microgrid are listed in Table \ref{table 2}. The incidence matrix for the topology of the AC microgrid is given by 
\begin{align}
\mathcal B =
\left[
\begin{array}{rrrr}
-1 & 0 & 0 & -1 \\
1 & -1 &  0 & 0 \\
0 & 1 & -1 & 0\\
0 & 0 & 1 & 1
\end{array}
\right].
\end{align}

The parameters for the human behavior models are given as follows, $A=0.5\times \mathbf I$, $C=\text{diag}(0.08\,\, 0.08\,\, 0.12\,\, 0.12)$,
$D =\text{diag}(0.12\,\, 0.12\,\, 0.08\,\, 0.08)$ and $\mathcal L =0.4 \times \mathcal B   \mathcal B ^ T$. Moreover, we let $p^{\text{ego}}=[0.9\,\, 0.9\,\, 0.9\,\, 0.9]^T$, $p^{\text{bio}}=[0.6 \, \, 0.6 \, \, 0.6 \, \, 0.6]^T$ and $p=[0.8\,\,0.8\,\, 0.8\,\, 0.8]^T$ at $t=0$.

\begin{figure}[t]
	\begin{center}
		\includegraphics[width=0.51  \textwidth]{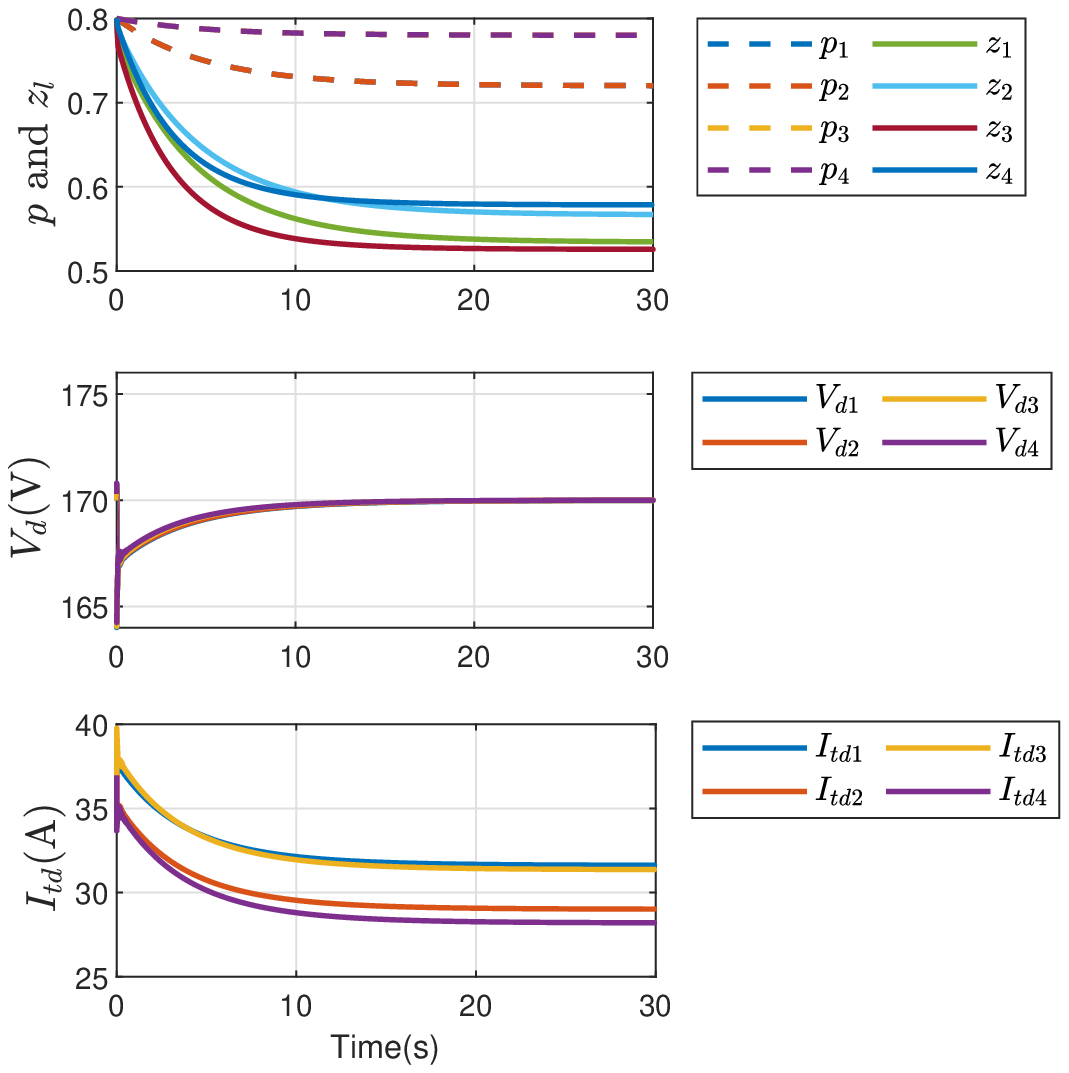}  \\
		\vspace{-6mm}
		\linespread{1}\caption{Case i) }  \label{Scenario 1}
		\vspace{4mm}
		\includegraphics[width=0.51  \textwidth]{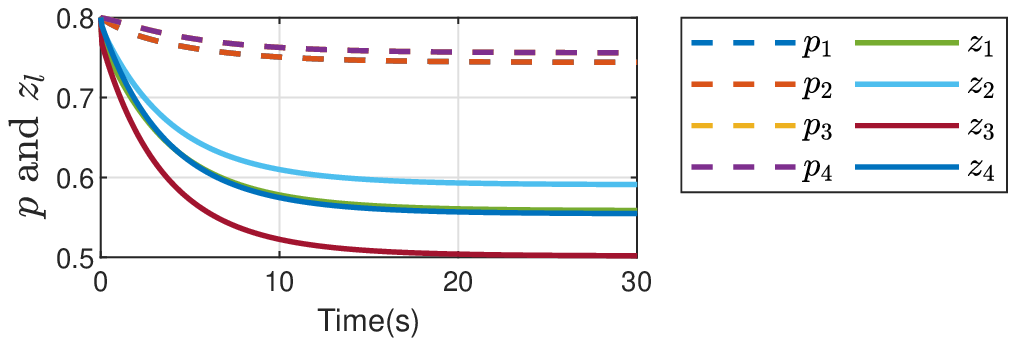}  \\
				\vspace{-2mm}
		\linespread{1}\caption{Case ii)  }  \label{Scenario 2}
	\end{center}
\end{figure}

\emph{Case i)} We first consider the case of HPS in (\ref{compact all}). The simulation results are presented in Figure \ref{Scenario 1}. In the first plot of Figure \ref{Scenario 1}, one can see that due to the influence of biospheric values $p^{\text{bio}}$,
$p$ indeed decreases. However, $p$ finally reaches the steady values 0.72 for $p_1$ and $p_2$ and 0.78 for $p_3$ and $p_4$ due to the different priorities to their egoistic and biospheric values. Prosumers 3 and 4 are more attached to their egoistic values with $c_i= 0.6$ compared with prosumers 1 and 2 with $c_i=0.4$, and hence are reluctant to change $p_i$ to a lower value depending on $p_i ^{\text{bio}}=0.6$. 
In view of (\ref{steady human ac}), due to the impact of incentives $s$, we know that $\bar z_l \le \bar p$. Indeed, in this example, we have
$\bar z_l=[  
0.53\,\,	0.57\,\,	0.53\,\,	0.58]^T < \bar p$ with  $\bar s = [ 
0.21\,\,	0.17\,\,	0.28\,\,	0.22]^T$.
Recalling the selected $A$, $C$ and $D$ in this example, the first plot in Figure \ref{Scenario 1} also shows the relatively fast transient process of $z_l$ compared with $p$, which demonstrates the fast and slow transient processes of behavior and personal norms, respectively.

For voltage regulation, from the second plot in Figure \ref{Scenario 1}, one can see that $V_d$ converges to the values around $V_r$. We also achieve $V_q \to \mathbf 0$, and we omit the plot due to space limitation. Now we consider the status of current sharing. From the third plot in Figure \ref{Scenario 1}, one can see that the prosumers
approximately share the total demand due to $\pi_{ci}=1$ for all $i$ (the generated currents $I_{tdi}$ are similar each other). Differently, if we let for instance $\pi_{c4}=100$ and $\pi_{ci}=1$ ($i=1, 2, 3$), which implies that prosumer 4 has a much smaller generation capacity, then the steady current generation would be $\bar I_{td}= [35\,\,	27\,\,	35\,\,	4.2]^T$.


\emph{Case ii)} The simulation results for the human behavior model in (\ref{compact human 1})  including social influence are shown in Figure \ref{Scenario 2}. 
Since the Laplacian matrix $\mathcal L$ in (\ref{compact human 1}) in this example represents a connected graph, one can see that the components in $p$ finally converge values that are closer to each other with respect to case i), despite they have different weights on egoistic and biospheric values.  
Furthermore, under the obtained incentives $\bar s = [0.21 \,\, 0.17 \,\, 0.28 \,\, 0.22 ]^T$ (which does not change in view of case i)), $z_l$ converges to $[0.56\,\,0.59 \,\, 0.50 \,\,0.55 ] ^ T < \bar p$.

Finally, the total consumption reduction can be calculated by $\mathbf 1^T I_{Ld}(\mathbf 1 - \bar z_l) $, and the amounts are 49.87 A (44.93\%) and 50.15 A (45.18 \%), respectively.

%
%


\section{Conclusions and future research}
 In this paper, we formulate a framework to bridge the disciplines of systems \& control and environmental psychology. Specifically, we formulate a HPS framework to describe energy saving behavior of humans in social networks and also their interactions with an AC power grid. It is clear that the developed second-order models describing human energy saving behavior and personal norms are consistent with the findings in environmental psychology and able to differentiate an individual's extrinsic and intrinsic drivers of energy saving behavior, and also allow to model their fast and slow transient processes, respectively. 
 With the developed HPS, we formulate a social-physical welfare optimization problem and provide the design of primal-dual controller. It is proved that the controller computing optimal incentives to humans and control inputs to an AC power grid can 
stabilize the closed-loop system whose state converges to an equilibrium solving the optimization problem. 
 The developed models of human activities not only interpret the findings in psychology from the control viewpoint, but also are consistent with the studies on opinion dynamics.

The results in this paper can be extended towards various directions in the future. 
Among these, one possible direction is to collect social and electrical data to identify the parameters of the proposed Behavior-Personal-norm models. Another direction is to consider inequality constraints (e.g., by considering line congestions or subsidy upper bounds) in the social-physical welfare optimization problem.  

\bibliographystyle{IEEEtran}
\bibliography{ACbib1}

\begin{thebibliography}{10}
\providecommand{\url}[1]{#1}
\csname url@samestyle\endcsname
\providecommand{\newblock}{\relax}
\providecommand{\bibinfo}[2]{#2}
\providecommand{\BIBentrySTDinterwordspacing}{\spaceskip=0pt\relax}
\providecommand{\BIBentryALTinterwordstretchfactor}{4}
\providecommand{\BIBentryALTinterwordspacing}{\spaceskip=\fontdimen2\font plus
\BIBentryALTinterwordstretchfactor\fontdimen3\font minus
  \fontdimen4\font\relax}
\providecommand{\BIBforeignlanguage}[2]{{%
\expandafter\ifx\csname l@#1\endcsname\relax
\typeout{** WARNING: IEEEtran.bst: No hyphenation pattern has been}%
\typeout{** loaded for the language `#1'. Using the pattern for}%
\typeout{** the default language instead.}%
\else
\language=\csname l@#1\endcsname
\fi
#2}}
\providecommand{\BIBdecl}{\relax}
\BIBdecl

\bibitem{steg2015understanding}
L.~Steg, G.~Perlaviciute, and E.~van~der Werff, ``Understanding the human
  dimensions of a sustainable energy transition,'' \emph{Frontiers in
  psychology}, vol.~6, p. 805, 2015.

\bibitem{stern2016opportunities}
P.~C. Stern, K.~B. Janda, M.~A. Brown, L.~Steg, E.~L. Vine, and L.~Lutzenhiser,
  ``Opportunities and insights for reducing fossil fuel consumption by
  households and organizations,'' \emph{Nature Energy}, vol.~1, no.~5, pp.
  1--6, 2016.

\bibitem{stern2016towards}
P.~C. Stern, B.~K. Sovacool, and T.~Dietz, ``Towards a science of climate and
  energy choices,'' \emph{Nature Climate Change}, vol.~6, no.~6, pp. 547--555,
  2016.

\bibitem{sovacool2014diversity}
B.~K. Sovacool, ``Diversity: energy studies need social science,'' \emph{Nature
  News}, vol. 511, no. 7511, p. 529, 2014.

\bibitem{schwartz1981normative}
S.~H. Schwartz and J.~A. Howard, ``A normative decision-making model of
  altruism,'' \emph{Altruism and helping behavior}, pp. 189--211, 1981.

\bibitem{stren2000toward}
P.~Stren, ``Toward a coherent theory of environmentally significant
  behaviour,'' \emph{Journal of Social Issues}, vol.~56, no.~3, pp. 407--424,
  2000.

\bibitem{stern1999value}
P.~C. Stern, T.~Dietz, T.~Abel, G.~A. Guagnano, and L.~Kalof, ``A
  value-belief-norm theory of support for social movements: The case of
  environmentalism,'' \emph{Human ecology review}, pp. 81--97, 1999.

\bibitem{dietz2005environmental}
T.~Dietz, A.~Fitzgerald, and R.~Shwom, ``Environmental values,'' \emph{Annu.
  Rev. Environ. Resour.}, vol.~30, pp. 335--372, 2005.

\bibitem{steg2018drives}
L.~Steg, R.~Shwom, and T.~Dietz, ``What drives energy consumers?: Engaging
  people in a sustainable energy transition,'' \emph{IEEE Power and Energy
  Magazine}, vol.~16, no.~1, pp. 20--28, 2018.

\bibitem{schwartz2015advertising}
D.~Schwartz, W.~Bruine~de Bruin, B.~Fischhoff, and L.~Lave, ``Advertising
  energy saving programs: The potential environmental cost of emphasizing
  monetary savings.'' \emph{Journal of Experimental Psychology: Applied},
  vol.~21, no.~2, p. 158, 2015.

\bibitem{nolan2008normative}
J.~M. Nolan, P.~W. Schultz, R.~B. Cialdini, N.~J. Goldstein, and
  V.~Griskevicius, ``Normative social influence is underdetected,''
  \emph{Personality and social psychology bulletin}, vol.~34, no.~7, pp.
  913--923, 2008.

\bibitem{schultz2007constructive}
P.~W. Schultz, J.~M. Nolan, R.~B. Cialdini, N.~J. Goldstein, and
  V.~Griskevicius, ``The constructive, destructive, and reconstructive power of
  social norms,'' \emph{Psychological science}, vol.~18, no.~5, pp. 429--434,
  2007.

\bibitem{degroot1974reaching}
M.~H. DeGroot, ``Reaching a consensus,'' \emph{Journal of the American
  Statistical Association}, vol.~69, no. 345, pp. 118--121, 1974.

\bibitem{friedkin1990social}
N.~E. Friedkin and E.~C. Johnsen, ``Social influence and opinions,''
  \emph{Journal of Mathematical Sociology}, vol.~15, no. 3-4, pp. 193--206,
  1990.

\bibitem{ye2020continuous}
M.~Ye, M.~H. Trinh, Y.-H. Lim, B.~D. Anderson, and H.-S. Ahn, ``Continuous-time
  opinion dynamics on multiple interdependent topics,'' \emph{Automatica}, vol.
  115, p. 108884, 2020.

\bibitem{wongkaew2015control}
S.~Wongkaew, M.~Caponigro, and A.~Borzi, ``On the control through leadership of
  the hegselmann--krause opinion formation model,'' \emph{Mathematical Models
  and Methods in Applied Sciences}, vol.~25, no.~03, pp. 565--585, 2015.

\bibitem{caponigro2016sparse}
M.~Caponigro, B.~Piccoli, F.~Rossi, and E.~Tr{\'e}lat, ``Sparse feedback
  stabilization of multi-agent dynamics,'' in \emph{IEEE Conference on Decision
  and Control}, 2016, pp. 4278--4283.

\bibitem{amelkin2017polar}
V.~Amelkin, F.~Bullo, and A.~K. Singh, ``Polar opinion dynamics in social
  networks,'' \emph{IEEE Transactions on Automatic Control}, vol.~62, no.~11,
  pp. 5650--5665, 2017.

\bibitem{nasirian2014distributed}
V.~Nasirian, S.~Moayedi, A.~Davoudi, and F.~L. Lewis, ``Distributed cooperative
  control of dc microgrids,'' \emph{IEEE Transactions on Power Electronics},
  vol.~30, no.~4, pp. 2288--2303, 2014.

\bibitem{de2018power}
C.~De~Persis, E.~R. Weitenberg, and F.~D{\"o}rfler, ``A power consensus
  algorithm for dc microgrids,'' \emph{Automatica}, vol.~89, pp. 364--375,
  2018.

\bibitem{cucuzzella2018robust}
M.~Cucuzzella, S.~Trip, C.~De~Persis, X.~Cheng, A.~Ferrara, and A.~van~der
  Schaft, ``A robust consensus algorithm for current sharing and voltage
  regulation in dc microgrids,'' \emph{IEEE Transactions on Control Systems
  Technology}, vol.~27, no.~4, pp. 1583--1595, 2018.

\bibitem{trip2018distributed}
S.~Trip, M.~Cucuzzella, X.~Cheng, and J.~Scherpen, ``Distributed averaging
  control for voltage regulation and current sharing in dc microgrids,''
  \emph{IEEE Control Systems Letters}, vol.~3, no.~1, pp. 174--179, 2018.

\bibitem{cucuzzella2019distributed}
M.~Cucuzzella, K.~C. Kosaraju, T.~Bouman, G.~Schuitema, S.~Johnson-Zawadzki,
  C.~Fischione, L.~Steg, and J.~Scherpen, ``Distributed control of dc grids: a
  social perspective,'' \emph{arXiv preprint arXiv:1912.07341}, 2019.

\bibitem{stegink2016unifying}
T.~Stegink, C.~De~Persis, and A.~van~der Schaft, ``A unifying energy-based
  approach to stability of power grids with market dynamics,'' \emph{IEEE
  Transactions on Automatic Control}, vol.~62, no.~6, pp. 2612--2622, 2016.

\bibitem{kosaraju2019distributed}
K.~C. Kosaraju, M.~Cucuzzella, and J.~M. Scherpen, ``Distributed control of dc
  microgrids using primal-dual dynamics,'' in \emph{IEEE Conference on Decision
  and Control}.\hskip 1em plus 0.5em minus 0.4em\relax IEEE, 2019, pp.
  6215--6220.

\bibitem{zhao2014design}
C.~Zhao, U.~Topcu, N.~Li, and S.~Low, ``Design and stability of load-side
  primary frequency control in power systems,'' \emph{IEEE Transactions on
  Automatic Control}, vol.~59, no.~5, pp. 1177--1189, 2014.

\bibitem{guerrero2007decentralized}
J.~M. Guerrero, J.~Matas, L.~G. de~Vicuna, M.~Castilla, and J.~Miret,
  ``Decentralized control for parallel operation of distributed generation
  inverters using resistive output impedance,'' \emph{IEEE Transactions on
  industrial electronics}, vol.~54, no.~2, pp. 994--1004, 2007.

\bibitem{golsorkhi2016gps}
M.~S. Golsorkhi, M.~Savaghebi, D.~D.-C. Lu, J.~M. Guerrero, and J.~C. Vasquez,
  ``A gps-based control framework for accurate current sharing and power
  quality improvement in microgrids,'' \emph{IEEE Transactions on Power
  Electronics}, vol.~32, no.~7, pp. 5675--5687, 2016.

\bibitem{park1929two}
R.~H. Park, ``Two-reaction theory of synchronous machines generalized method of
  analysis-part i,'' \emph{Transactions of the American Institute of Electrical
  Engineers}, vol.~48, no.~3, pp. 716--727, 1929.

\bibitem{cucuzzella2018cooperative}
M.~Cucuzzella, S.~Trip, A.~Ferrara, and J.~Scherpen, ``Cooperative voltage
  control in {AC} microgrids,'' in \emph{IEEE Conference on Decision and
  Control}, 2018, pp. 6723--6728.

\bibitem{bouman2019motivating}
T.~Bouman and L.~Steg, ``Motivating society-wide pro-environmental change,''
  \emph{One Earth}, vol.~1, no.~1, pp. 27--30, 2019.

\bibitem{bouman2018measuring}
T.~Bouman, L.~Steg, and H.~A. Kiers, ``Measuring values in environmental
  research: a test of an environmental portrait value questionnaire,''
  \emph{Frontiers in psychology}, vol.~9, p. 564, 2018.

\bibitem{schwartz2012refining}
S.~H. Schwartz, J.~Cieciuch, M.~Vecchione, E.~Davidov, R.~Fischer,
  C.~Beierlein, A.~Ramos, M.~Verkasalo, J.-E. L{\"o}nnqvist, K.~Demirutku
  \emph{et~al.}, ``Refining the theory of basic individual values.''
  \emph{Journal of personality and social psychology}, vol. 103, no.~4, p. 663,
  2012.

\bibitem{steg2016values}
L.~Steg, ``Values, norms, and intrinsic motivation to act proenvironmentally,''
  \emph{Annual Review of Environment and Resources}, vol.~41, pp. 277--292,
  2016.

\bibitem{stern1994value}
P.~C. Stern and T.~Dietz, ``The value basis of environmental concern,''
  \emph{Journal of social issues}, vol.~50, no.~3, pp. 65--84, 1994.

\bibitem{dogan2014making}
E.~Dogan, J.~W. Bolderdijk, and L.~Steg, ``Making small numbers count:
  environmental and financial feedback in promoting eco-driving behaviours,''
  \emph{Journal of Consumer Policy}, vol.~37, no.~3, pp. 413--422, 2014.

\bibitem{bouman2020insights}
T.~Bouman, L.~Steg, and T.~Dietz, ``Insights from early covid-19 responses
  about promoting sustainable action,'' \emph{Nature Sustainability}, pp. 1--7,
  2020.

\bibitem{stern2000new}
P.~C. Stern, ``New environmental theories: toward a coherent theory of
  environmentally significant behavior,'' \emph{Journal of social issues},
  vol.~56, no.~3, pp. 407--424, 2000.

\bibitem{boyd2004convex}
S.~Boyd, S.~P. Boyd, and L.~Vandenberghe, \emph{Convex optimization}.\hskip 1em
  plus 0.5em minus 0.4em\relax Cambridge university press, 2004.

\bibitem{kosaraju2020differentiation}
K.~C. Kosaraju, M.~Cucuzzella, J.~M. Scherpen, and R.~Pasumarthy,
  ``Differentiation and passivity for control of {B}rayton-{M}oser systems,''
  \emph{IEEE Transactions on Automatic Control}, 2020.

\bibitem{hardy1952inequalities}
G.~H. Hardy, J.~E. Littlewood, G.~P{\'o}lya, G.~P{\'o}lya, D.~Littlewood
  \emph{et~al.}, \emph{Inequalities}.\hskip 1em plus 0.5em minus 0.4em\relax
  Cambridge university press, 1952.

\end{thebibliography}

\end{document}